\UseRawInputEncoding
\documentclass[%
 reprint,
 amsmath,amssymb,
 aps,
]{revtex4-2}

\usepackage{graphicx}
\usepackage{dcolumn}
\usepackage{bm}

    \usepackage[title]{appendix}
    \usepackage{comment}
    \usepackage{xcolor}
    \definecolor{darkblue}{RGB}{0,0,50.2}
    \definecolor{royalblue}{RGB}{17,30,108}
    \definecolor{aquamrine}{RGB}{127,255,212}
    \definecolor{turquoise}{RGB}{64,224,208}
    \definecolor{steelblue}{RGB}{70,130,180}
    \definecolor{mygreen}{RGB}{28,172,0}
    \definecolor{mylilas}{RGB}{170,55,241}
    \usepackage{hyperref}
    \hypersetup{
        colorlinks=true,
        linkcolor=steelblue,
        filecolor=magenta,      
        urlcolor=steelblue,
        citecolor=mylilas,
        pdfpagemode=FullScreen,
        }
    \usepackage{epsfig}
    \usepackage{float}
\begin{document}


\title{The Ground-state Inter-plane Superconducting Coherence Length of La$_{1.875}$Sr$_{0.125}$CuO$_4$ Measured by a ``Xiometer"}

\author{Itay Mangel}
    \email{itaymangel89@campus.technion.ac.il}
\author{Amit Keren}
    \email{phkeren@technion.ac.il}
\affiliation{Department of Physics, Technion-Israel Institute of Technology, Haifa, 3200003, Israel}

\date{\today} 

\begin{abstract}
A long excitation coil piercing a superconducting (SC) ring is used to generate ever increasing persistent current in the ring, until the current destroys the order parameter. Given that the penetration depth $\lambda$ is known, this experiment measures, hypothetically, the coherence length $\xi$, hence the name "Xiometer". We examine various aspects of this theoretically driven hypothesis by testing niobium rings with different dimensions, and by comparing the results to the known values of $\xi$. We then apply the method to two La$_{1.875}$Sr$_{0.125}$CuO$_4$ rings at $T \rightarrow 0$. In one, the current flows in the CuO$_2$ planes hence it is set by $\xi_{ab}$. In the other, the current must cross planes and is determined by $\xi_{c}$. We find that $\xi_{c}=1.3 \pm 0.1$~nm, and $\xi_{ab}<2.3~$nm indicating that at low temperatures the Cooper pairs are three dimensional.
\end{abstract}

\maketitle

In the world of superconductivity there are two important length scales: the penetration depth $\lambda$, and the coherence length $\xi$. From an application point of view, $\lambda^{-2}$ is a measure of superconducting carrier density and determines the current carrying capabilities of a superconductor, while $\xi$ sets the dimensions of Josephson junctions. There is an arsenal of experimental methods to directly measure the superconducting stiffness $\rho_{s}$ and find $\lambda$ via the relation $\rho_{s} = \frac{1}{\mu_{0}\lambda^2}$ in different crystal orientations. However, methods of measuring $\xi$ are limited. In the cuprates the coherence length in the $ab$ plane ($\xi_{ab}$) was measured by: electron–boson spectral density function \cite{Hwang123SuperconductingFunction} and vortex diameter via scanning tunneling microscopy (STM) \cite{Pan2000STM81d}. More commonly, one finds $H_{c2}$ near $T_c$ using one of various methods: resistivity \cite{Oh1988UpperX}  vortex-Nernst effect \cite{Wang2003DependenceCuprates}, specific heat \cite{Wang2008DopingHeat}, or thermal conductivity \cite{Grissonnanche2014DirectSuperconductors}, extrapolating to $T=0$ using theories that are not necessarily accurate over the whole temperature range, and uses the relation $\xi=\sqrt{\Phi_0/2 \pi H_{c2}  }$ \cite{Oh1988UpperX, Petrenko2022TemperatureFilms}. All methods find $\xi_{ab}$ on the scale of $1.5-3$~nm.

Measuring coherence length in the $c$ direction ($\xi_c$) in cuprates is more difficult since it is smaller and because cleaving in a plane including the $c$ axis is challenging, making scanning techniques nearly impossible. $\xi_c$ is bound by $1.5$~nm, the thickness of a superconducting Bi$_2$Sr$_2$CaCu$_2$O$_{8+x}$ (Bi2212) monolayer \cite{Yu2019High-temperatureBi2Sr2CaCu2O8+}. Extrapolations from high temperatures lead to $\xi \sim 1$~\AA ~\cite{Petrenko2022TemperatureFilms,Oh1988UpperX}. The only measured value of $\xi_c=0.86$~nm at $T \rightarrow 0$, as far as we know, is from an $H_{c2}$ of $250$~T obtained using the electromagnetic flux compression method \cite{SekitaniUpper7Ad}. This number has been questioned due to the transient nature of the magnetic field. Exact determination of $\xi_c$ is becoming exceedingly important due to Josephson junctions created by twisted Bi2212 crystals \cite{Lee2022EncapsulatingDisorder,Klemm2005TheImplications,Song2022DopingCuprates,Lee2021TwistedSuperconductor,Takano2002D-likeJunctions} showing fractional Shapiro steps \cite{Zhao2021EmergentSuperconductors,Tummuru2022JosephsonBilayers},
and as a challenge for cuprates theory.

A new approach for measuring $\xi$ was suggested, and a very simple analysis formula was given, in Ref.~[\onlinecite{Gavish2021GinzburgLandauDevice}]. We name this approach ``Xiometer''. Here we briefly present the approach, justify the formula intuitively, and test it on Nb. Then, we apply it to $\xi$ measurements in La$_{1.875}$Sr$_{0.125}$CuO$_4$ (LSCO-1/8). We find that the $\xi_c=1.3 \pm 0.1$~nm determined by the Xiometer at $T \rightarrow 0$ is similar to the one obtained from $H_{c2}=250$~T of Ref.~\cite{SekitaniUpper7Ad} and calculation in Ref.~\cite{Petrenko2022TemperatureFilms}. The implication of this finding is that the Cooper pairs are more spherical than previously thought.

\begin{figure}
    \centering
    \includegraphics[width=\linewidth]{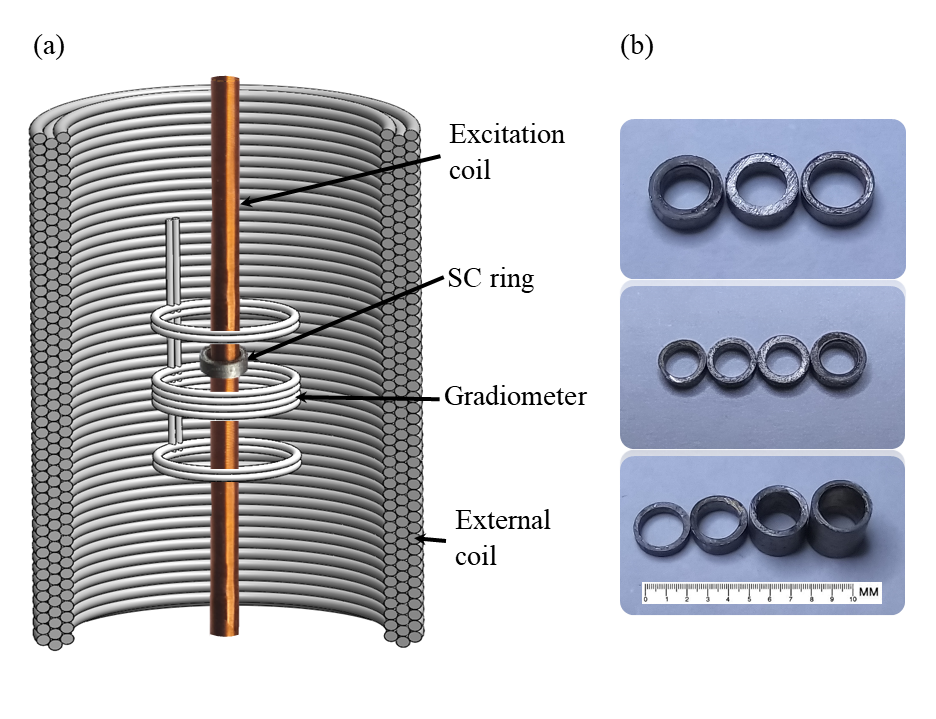}
    \caption{ {\bf Experimental setup:} (a) A niobium ring on a SC excitation coil (photo), an illustration of the  gradiometer, and the external coil that serves as a shim. (b) Niobium rings with different inner radii (up), outer radii (middle), and different height (down).
    }
    \label{fig:niobium_ring_and_experimental_setup}
\end{figure}

The Xiometer, like the Stiffnessometer \cite{Mangel2020Stiffnessometer:Application}, is based on a long, current carrying, excitation coil (EC) piercing a superconducting (SC) ring and a measurement of the ring's magnetic moment. The measurement is done by moving both ring and EC rigidly relative to a gradiometer type pick-up loops as demonstrated in Fig.~\ref{fig:niobium_ring_and_experimental_setup}(a). The gradiometer is connected to a superconducting quantum interference device (SQUID). The ring and EC are cooled to $T<T_c$, only then the current in the EC is turned on, and the magnetic moment of the ring is measured. The gauge invariant London equation states that

\begin{equation}
\boldsymbol{j} = -\rho_{s}\left( \boldsymbol{A}_{tot} - \frac{\Phi_0}{2\pi}\boldsymbol{\nabla} \varphi \right)
\label{eq: London}
\end{equation}
where $\boldsymbol{j}$ is the superconducting current density, $\boldsymbol{A}_{tot}$ is the total vector potential, $\Phi_0$ is the flux quanta, and $\varphi$ is the phase of the superconductor's order parameter. $\boldsymbol{A}_{tot}$ is a combination of the EC vector potential $\boldsymbol{A}_{ec} = \frac{\Phi_{ec}}{2 \pi r}\hat{\varphi}$, and the SC ring self induced vector potential $\boldsymbol{A}_{sc}$. 

To minimize the kinetic energy after cooling, the superconductor sets its own phase gradient to zero. Since this phase is quantized, turning the EC current on, gently, conserve the phase. Therefore, at the start of a measurement, the London equation $\boldsymbol{j} = -\rho_{s} \boldsymbol{A}_{tot} $ is valid. In principle, this relation can be used to determine the stiffness and in this case the apparatus works as a Stiffnessometer. However, for bulk crystals at low temperatures, when $\lambda$ is mach smaller than all dimensions of the sample, the SC ring exactly expels the applied flux, to keep $A_{tot}=0$ deep inside the ring, regardless of $\lambda$. In this case, the apparatus cannot be used to properly determine the stiffness, but only the break point of the London equation, which is set by $\lambda \xi$ as we explain below. For crystals at $T\ll T_c$ we use literature values of $\lambda$ and determine $\xi$. Hence, the name Xiometer.

The EC is home made from a NbTi SC wire with: $8$ layers of $600$ windings each ($4800$ in total), a wire diameter $0.106$~mm, core diameter $0.35$~mm, outer diameter $1.95$~mm, coil length $60$~mm, and flux to current ratio of $1.21 \cdot 10^{-7} ~ Tm^2/A$. The apparatus is an add-on to a Cryogenics S700X SQUID magnetometer. External magnetic fields can be canceled by an external coil shown in  Fig.~\ref{fig:niobium_ring_and_experimental_setup}(a), with a resolution of $5\times 10^{-7}$~T. The gradiometer radius is $R_{pl}=13$~mm, its total height is $14.0$~mm, and it is made of: two windings clockwise, four anticlockwise, and two clockwise also depicted in Fig.~\ref{fig:niobium_ring_and_experimental_setup}(a). The set of Nb rings with different dimensions used in the first part of this experiment is shown in Fig.~\ref{fig:niobium_ring_and_experimental_setup}(b).

\begin{figure}
    \centering
    \includegraphics[width=\linewidth]{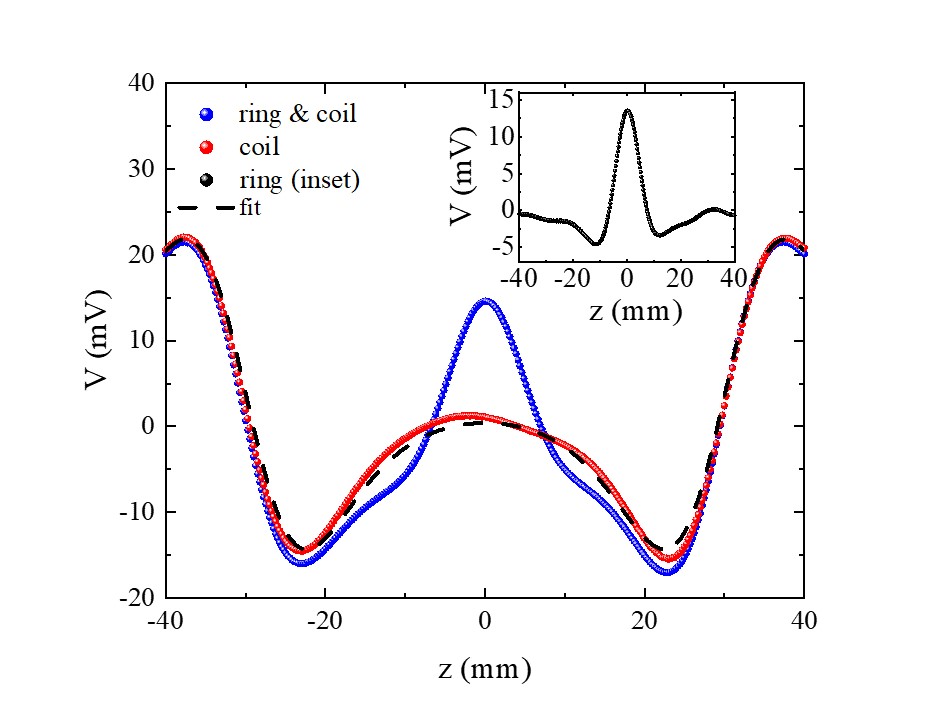}
    \caption{{\bf Raw data.} Main: The squid output voltage $V$ as a function of the relative position $z$ of a Nb ring and coil to the gradiometer center. The EC current is $10.0$~mA, $T=1.6$~K, and the ring dimensions are: $r_{in} = 1.0$~mm, $r_{out} = 1.75$~mm, and $h=1.0$~mm. Red spheres is the EC signal without the ring. Blue spheres is the combined signal of the ring and coil. Dashed black line show a numerical fit used to determine the conversion factor between the output voltage of the SQUID and the magnetic moment of a sample. Inset: The subtraction of the two measurements giving the ring's signal. The data presented is after subtraction of a linear component.}
    \label{fig:raw_data_ring_and_coil}
\end{figure}

A typical data set of the SQUID output voltage $V$ as a function of ring position $z$, is depicted in Fig.~\ref{fig:raw_data_ring_and_coil} with and without a ring. In these measurements the EC was allowed to transverse the gradiometer from one side to the next. The peaks and valleys away from $z=0$ are due to the ends of the coil moving through the different winding groups of the gradiometer. Without the ring (red symbols) a moderately concave signal is observed around $z=0$. This occurs when the center of the gradiometer and the center of the coil are at the same height. In this situation, flux through the gradiometer due to the EC barely changes, therefore the measurement is sensitive mostly to $\boldsymbol{A}_{sc}$. With the ring, a new signal (blue symbols) appears around $z=0$. The difference is the net SC ring's signal (inset) and its amplitude is proportional to the magnetic moment $m$ (or $A_{sc}$) of the ring. We note that a linear base-line was subtracted from both data sets due to the EC asymmetry (wires enter and exit from one side only). Linear base-line subtraction is irrelevant for the data analysis (see below). 

\begin{figure}
    \centering
    \includegraphics[width=\linewidth]{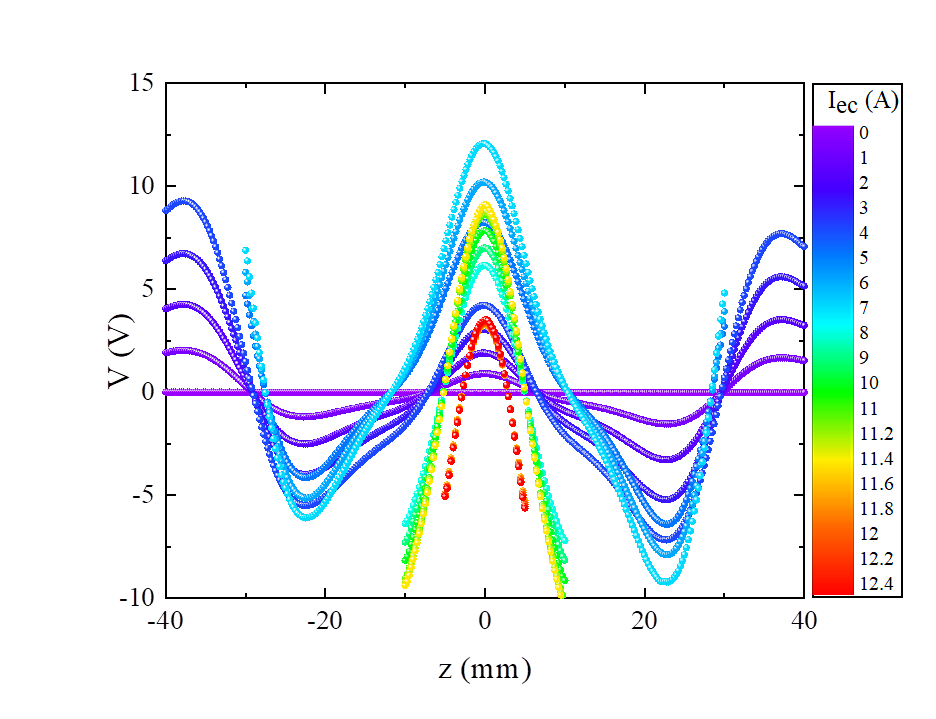}
    \caption{{\bf Raw data in a limited scan length.} SQUID output voltage as a function of the ring's position $z$ for different applied EC currents. For low currents ($0-4~A$) it is possible to measure the full signal of the coil and the ring (see Fig.~\ref{fig:raw_data_ring_and_coil}). For higher currents the coil's signal exceeds the detector's dynamic range and a measurement on a shorter range is needed. Consequently, the edges of the coil are not observed and the signal amplitude cannot be determined. Therefore, the second derivative of the signal $V''_0$ is used to determine the moment as explained in the text.}
    \label{fig:raw_data_currants}
\end{figure}

As the EC current increases, the signal from the EC traversing the gradiometer overwhelms the ring's signal, as is clear from Fig.~\ref{fig:raw_data_currants}. Therefore, we limit the motion to a small region around $z=0$. Consequently it is impossible to detect the bottom of the signal and evaluate its amplitude. However, it is clear that the top of the peak at $z=0$ becomes sharper with increasing current. Therefore, we use the second derivative of the SQUID's output voltage $V''_0=\frac{d^2V}{d^2z}(z=0)$ as a measure of $m$. This method also eliminates the undesired linear contribution of the coil's asymmetry. The conversion from $V''_0$ to $m$ is explained shortly.

Finally, to keep the leads, coil, and ring cooled, liquid helium is sprayed via a diffuser from the bottom of the sample chamber on the EC and sample, and pumped along the current leads all the way to the top of the cryostat just before thicker leads are connected to the power supply. This way the EC remains cold even when currents of more than $10$~A are applied.
    
The measured samples are $99.9\%$ pure Nb rings, and two LSCO-$1/8$ single crystal rings, grown, using a traveling solvent floating zone technique and cut with a laser cutter. Laue x-ray diffraction is used to identify the orientation of the crystals and the two rings are cut in different orientations of the CuO$_2$ planes relative to ring's symmetry axis. In one ring the CuO$_2$ planes are parallel to the symmetry axis. In the other ring the planes are perpendicular to this axis. We address the two rings as $a$ and $c$ rings, respectively. In the $c$ ring, current can flow around the ring on CuO$_2$ planes without crossing planes. In order to flow around the $a$ ring current must cross CuO$_2$ planes.

\begin{figure}
    \centering
    \includegraphics[width=\linewidth]{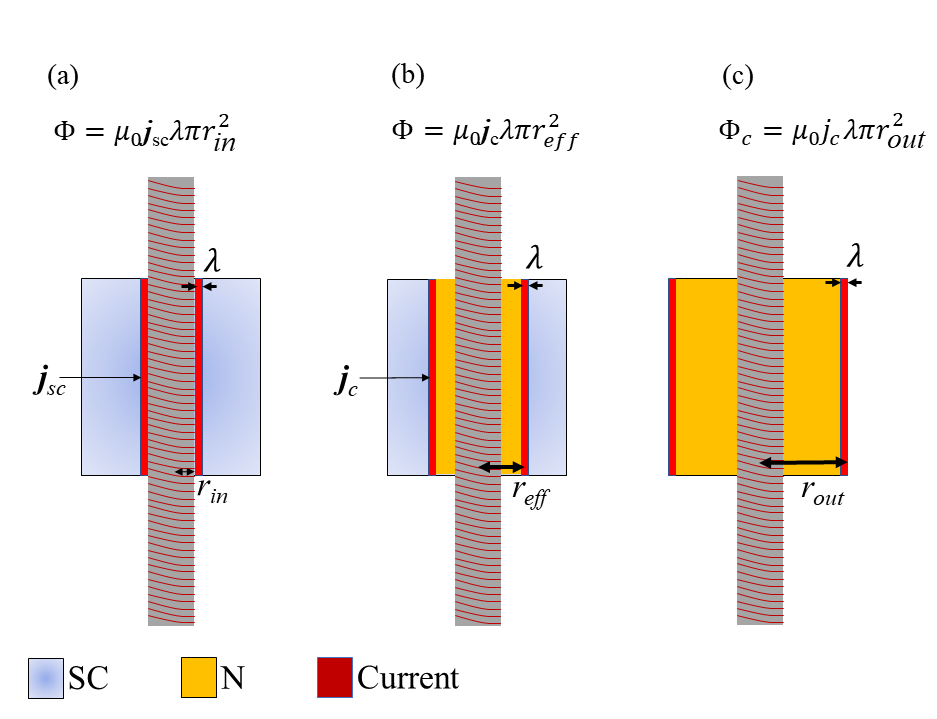}
    \caption{{\bf Schematic description of the superconductor screening currents evolution for increasing flux in the EC.} Each sketch shows cross sections of the EC and a cylindrical ring. Yellow and gray regions represent the normal and SC states, respectively. The current flows in the red region. The flux current relations are given at the top of each panel. (a) Low flux in the EC resulting in screening current along the inner rim of the cylinder. (b) Stronger flux in the EC forces the screening current to move outwards while the inner region of the cylinder becomes normal. (c) The critical flux is reached once the screening current reaches the outer radius of the cylinder and its bulk is no longer SC.}
    \label{fig:back_envelope_calculation}
\end{figure}

A detailed derivation of the Xiometer analysis theory can be found in Ref.~\cite{Gavish2021GinzburgLandauDevice}. Here we provide a back of the envelope calculation that gives the same answer up to a numerical factor. Since, as mentioned before, deep in the superconductor there are no currents (and no fields), $A_{tot}=0$. Therefore, the magnetic flux of the EC, $\Phi_{ec}$, is perfectly matched by the magnetic flux from the super-current $\Phi_{sc}$ in the ring, namely, $\Phi_{sc} =- \Phi_{ec}$. For low $\Phi_{ec}$ we assume that the supper current density $j$ is uniform along a cylinder of width $\lambda$ attached to the inner rim of the ring at $r_{in}$ as in Fig.~\ref{fig:back_envelope_calculation}(a). The magnetic flux of such a current is $\Phi_{sc} = \mu_0 \lambda j \pi r_{in}^2$. When $\Phi_{ec}$ increases, $j$ will also increases until it reaches the critical current density $j_{c}$ of the SC. When this happens the order parameter is destroyed next to the inner rim, and the current has to retreat from the inner rim to an effective radius $r_{eff}$ demonstrated in Fig.~\ref{fig:back_envelope_calculation}(b). This process continues until $r_{eff}=r_{out}$ as shown in Fig.~\ref{fig:back_envelope_calculation}(c). At this situation vortices start entering the sample, and the applied flux is named critical flux; it is given by  $\Phi_{c} = \mu_0 \lambda j_{c} \pi r_{out}^2$. Using the definition $j = e^{*}n v = \frac{m^{*}v}{\mu_{0} \lambda^2 e^{*}}$ where $n$, $v$, $e^{*}$, and $m^{*}$ are the carrier's density, velocity, charge, and mass respectively, and the relations of the critical momentum $m^{*}v_{c} = \frac{\hbar}{\sqrt{3}\xi}$ \cite{Tinkham2004Introduction1}, and the flux quanta $\Phi_0 =2\pi\hbar/e^{*} $ one finds
\begin{equation}
\frac{\Phi_{c}}{\Phi_0} = \frac{r_{out}^2}{2\sqrt{\alpha}\lambda\xi}
\label{eq:critical flux}
\end{equation}
where $\alpha=3$. In the exact derivation \cite{Gavish2021GinzburgLandauDevice} $\alpha=2$.

It should be pointed out that the SC produce a field in the volume where the order parameter is destroyed. One might wonder if this field penetrates as vortices into the SC when it is of Type-II. It was found in Ref.~\cite{Khanukov2022MixedField}, using a scanning SQUID, that in ultra thin film that show vortices due to sporadic magnetic field, the current in the coil does not add new vortices. This is not surprising since there is no pressure from twisted field lines bypassing the sample to penetrate into the sample. In our case the field lines are strait and in the center of the sample. They are in the most convenient place to be and do not need to penetrate the sample as vortices.

\begin{figure}
    \centering
    \includegraphics[width=\linewidth]{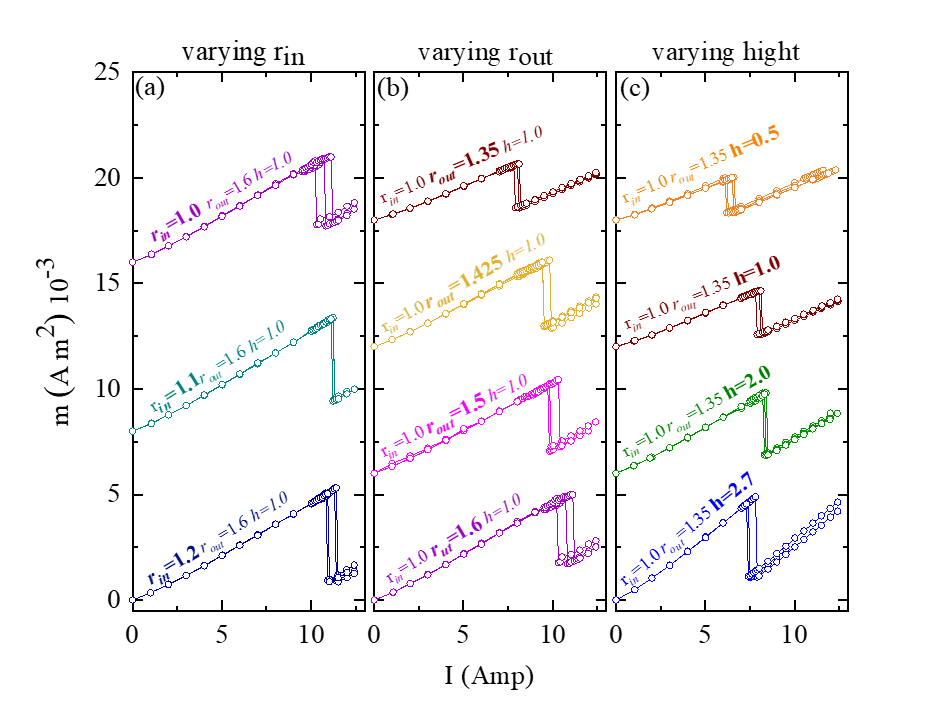}
    \caption{{\bf Critical current of different rings.} Nb ring's magnetic moment $m$ obtained using Eq.~\ref{eq:moment} as function of current in the excitation coil. The measurements are arranged in three sets. In each set only one parameter of the rings is changing. (a) Only the inner radius of the rings varies, (b) Only the outer radius of the rings differs, (c) Only the height of the rings changes. The experiment was done more then once with each ring for statistical purpose. The signal grow linearly with the EC current until it drops at some critical current $I_c$. The drop is due to a phase slip. The critical current varies between runs of the same ring due to thermal instability.}
    \label{fig:raw_data}
\end{figure}


The output voltage of the SQUID is related to the flux through the gradiometer by $V = K \Phi$ where $K$ is a conversion factor. When a sample with magnetic moment $m$ is located at height $z$ from the center of a single pickup-loop with radius $R_{pl}$, it's flux through the loop is $\Phi = \frac{\mu_0 R_{pl}^2 m}{2(R_{pl}^2 + z^2)^{3/2}}$. To calibrate $K$ we measured our coil for which we calculate the magnetic moment as a bundle of current loops with moment $m = \pi r^2 I$ each, where $r$ is the loop radius. The voltage output in this case is:

\begin{equation}
\begin{split}
& V(z) = \frac{\mu_0 R_{pl}^2 \pi I K}{2} \sum_{i} r_{i}^2 \left[  \frac{2}{( R_{pl}^2 + (z + z_{i} - \Delta_{pl})^2)^{3/2}} + \right. \\ &\left. \frac{2}{( R_{pl}^2 + (z + z_{i} + \Delta_{pl})^2)^{3/2}} - \frac{4}{( R_{pl}^2 + (z + z_{i})^2)^{3/2}} \right]
\label{eq:coil_gradiometer}
\end{split}
\end{equation}
where $z$ is the distance between the center of the coil and the center of the gradiometer, $\Delta_{pl}$ is the difference between the gradiometer's bundles, $r_i$ is the radius of the $i$'th layer, and $z_i$ the height of the $i$'th loop. By fitting Eq.~\ref{eq:coil_gradiometer} to a measurement of the coil with current of $10.0~mA$, as shown in Fig.~\ref{fig:raw_data_ring_and_coil}, we find the conversion factor to be $K=63.035$~V/Tmm$^{2}$. This procedure gives

\begin{equation}
m=4.95*10^{-10}\cdot V''_0
\label{eq:moment}
\end{equation}
where $V''_0$ is in units of mV/mm$^2$, and $m$ in units of Am$^2$.

To test Eq.~\ref{eq:critical flux} we measure a set of Nb rings with different inner and outer radii, and different height. Fig.~\ref{fig:raw_data} presents $m$ as a function of the applied EC current $I$. $m(I)$ is linear for low currents. At some high current $I_c$, a jump in $m$ is observed indicating that the critical flux in the coil $\Phi_c$ is reached, and that a phase slip has taken place. This process was done more than once for each ring for better statistics.

\begin{figure}
    \centering
    \includegraphics[width=\linewidth]{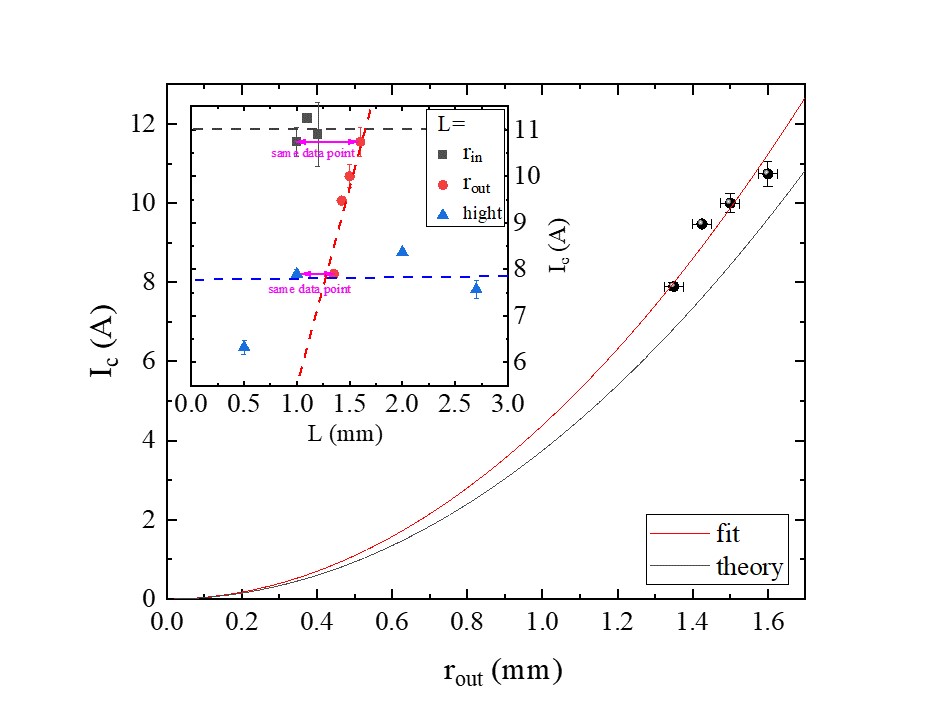}
    \caption{{\bf Critical current as function of different Nb ring sets.} Main, the set with changing outer radius. The black line follows Eq.~\ref{eq:critical flux} for $\lambda = 38$~nm and $\xi = 39$~nm \cite{Maxfield1965SuperconductingNiobium}. The red line is a parabolic fit to the data points (through the origin). The fit parameter mounts to $\lambda \xi=1267 \pm 32.5$~nm$^2$. Inset: black squares - rings with different inner radius, red circles - rings with varying outer radius, blue triangles - rings with changing height. The dashed lines are a guide to the eye.}
    \label{fig:all_parameters_and_r_out}
\end{figure}

The results of the experiment with the Nb rings is separated to three sets. In each set only one parameter is changing: inner radius $r_{in}$ panel (a), outer radius $r_{out}$ panel (b), and the ring's height $h$ panel (c). Variation of $r_{in}$ is limited because of the coil. Nevertheless, it does not seem to impact the $\Phi_c$. Variation of $r_{out}$ has noticeable and systematic influence on the critical flux as expected from Eq.~\ref{eq:critical flux}. Finally, between the smallest $h=0.5$~mm and all other values of $h$ a variation in $\Phi_c$ is detected. This is not expected from Eq.~\ref{eq:critical flux}. We ascribe this exception to the fact that Eq.~\ref{eq:critical flux} is derived in the limit of a tall, cylinder like, ring and the $h=0.5$~mm is not in this limit. A summary of $I_c$ as a function of $L$ ($L = r_{in} , r_{out} , h$) is depicted in the inset of Fig.~\ref{fig:all_parameters_and_r_out} where it is clear that $r_{out}$ is the most important parameter. The error bars are statistical.

In Fig.~\ref{fig:all_parameters_and_r_out} we present the $r_{out}$ dependence of $I_c$ on a full scale including the origin. 
When taking $\lambda$ and $\xi$ of Nb as $38$~nm and $39$~nm respectively from Ref. [\onlinecite{Maxfield1965SuperconductingNiobium}] and applying those in Eq.~\ref{eq:critical flux} (black line), we find reasonable agreement
between the measurements and theory. When fitting Eq.~\ref{eq:critical flux} to the data (red line), we find $\lambda \xi =1267 \pm 32.5$ nm$^2$ while the literature value is $1482$~nm$^2$.

\begin{figure}
    \centering
    \includegraphics[width=\linewidth]{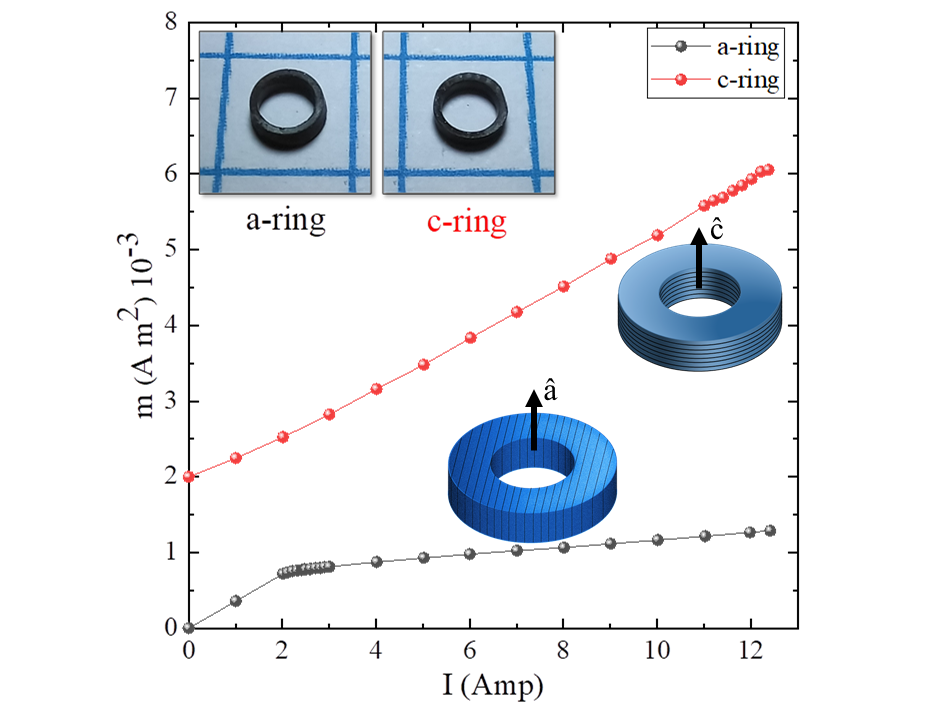}
    \caption{{\bf Critical flux determination of LSCO-$1/8$.} $m(I)$ for two rings. The insets show a pictures of the rings and the orientation of the planes. Their parameters are $r^a_{in}=1.05$~mm, $r^a_{out}=1.42$~mm, and $h_{a}=1.02$~mm, and $r^c_{in}=1.05$~mm, $r^c_{out}=1.35$~mm, and $h^{c}=0.89$~mm. $I_c$ of the a-ring is $2.2$~A and $I_c$ of the c-ring is higher than the maximum available current of $12.4$~A. }
    \label{fig:LSCO_and_rings}
\end{figure}

Having established Eq.~\ref{eq:critical flux} we use it to measure coherence length of the two LSCO-$1/8$ rings shown in the insets of Fig.~\ref{fig:LSCO_and_rings} with orientation and planes illustrated in the main panel. Previously, it was found that the persistent current critical temperature for the c-ring is higher than for the a-ring \cite{Kapon2019PhaseViewpoint}. Figure.~\ref{fig:LSCO_and_rings} shows $m$ at $T=1.6$~K as function of EC current for both the a-, and c-rings. At low currents, $m(I)$ is linear for both rings. But at about $2.2$~A, the a-ring moment $m_a$ has a break point where its behaviour changes. The c-ring moment $m_c(I)$ stays linear all the way up to our maximum current of $12.4$~A. We interpret this break point as the critical flux where vortices start entering the sample.
[floatfix]
At the experiment's temperature, the penetration depths of LSCO with $x=0.125$ are: $\lambda_{c}=4500$~nm [\onlinecite{Panagopoulos2000C-axisSuperconductors}], and $\lambda_{ab}=350$~nm ~[\onlinecite{Kapon2019PhaseViewpoint}]. The outer radii of the a- and c-rings are $r^a_{out}=1.42$~mm, and $r^c_{out}=1.35$~mm. Following Eq.~\ref{eq:critical flux} with $\lambda_{c}$ we find $\xi_{c}=1.3 \pm 0.1$~nm. The assumption here is that the bottleneck for current in the a-ring is the flow between planes (c-direction), and that the order parameter is destroyed first on the planes perpendicular to the flow. We can also place an upper bound on $\xi_{ab}<2.3$~nm using $\lambda_{ab}$.

The Ginzburg–Landau $\xi$ at $T=0$, which we measure with the Xiometer, is related to the Cooper pair size $\xi_0$ by a factor of 0.74 \cite{Tinkham2004Introduction1}. $\xi_0$, in turn, is set by $\hbar v_F/\Delta$ where $v_F$ is the Fermi velocity and delta is the superconducting gap. However, in the cuprates $\Delta$ varies along the Fermi surface, $v_F$ in the $c$ direction is not known, and there is no theory that can be used to extract more fundamental properties or be contrasted with our findings. A derivation of such a theory could be useful. 

To summarize, the relation between the dimensions of a superconducting ring, pierced by a long coil, and the critical flux in the coil, is tested. It is demonstrated that if the ring's height is similar to or bigger than its radii, only the outer radius of the ring is relevant, and the critical flux depends quadratically on this radius. Using this observation we measured $\xi_{ab}$ and $\xi_{c}$ of LSCO-$1/8$ at $T=1.6$~K. Despite the cuprates being very unisotropic systems, we found that $\xi_{c}$ is similar to the literature value of $\xi_{ab}$, indicating a 3D Cooper pair.

\section*{Acknowledgements} \label{sec:acknowledgements}
    This research was funded by the Israeli Science foundation grant number 3875/21.


%

\appendix*

\end{document}